\def\bar{\overline}

\def\e{\epsilon}

\def\bar{\overline}

\def\eV{{\rm eV}}
\def\ue3{|U_{e3}|}
\def\mnu{{\mathcal M}_{\nu }}
\def\be{\begin{equation}}
\def\ee{\end{equation}}

\def\ord{{\mathcal O}}

\def\ba{\begin{array}}
\def\ea{\end{array}}
\def\beqa{\begin{eqnarray}}
\def\eeqa{\end{eqnarray}}
\documentstyle [a4,epsfig,12pt]{article}
\begin{document}
\baselineskip=22 pt
\setcounter{page}{1}
\thispagestyle{empty}
\topskip 0.5  cm
\centerline{\large \bf Universal 2-3  Symmetry}
\vskip 1.5 cm
\centerline{{\large Anjan S. Joshipura}
\renewcommand{\thefootnote}{\fnsymbol{footnote}}
\footnote[3]{E-mail address: anjan@prl.res.in}
 }
\vskip 0.8 cm
 \centerline{ \it{Physical Research Laboratory, Navarangpura,
 Ahmedabad, 380 009, INDIA}}
\vskip 1.5 cm
\centerline{\bf ABSTRACT}\par
\vskip 0.8 cm

Possible maximal mixing seen in the oscillations of the atmospheric neutrinos
has led to postulate of
a $\mu$-$\tau$ symmetry which interchanges $\nu_\mu$ and $\nu_\tau$.
We argue that such symmetry need not be special to neutrinos
but can be extended to all fermions. The assumption that all fermion mass
matrices are approximately invariant under interchange of the second and
the third
generation fields is shown to be phenomenologically viable and has
interesting consequences. In the quark sector,
the smallness of $V_{ub}$ and $V_{cb}$ can be a consequences of this
approximate 2-3 symmetry.
The same approximate symmetry can simultaneously lead to large
atmospheric mixing angle and can describe the leptonic mixing quite
well provided the neutrino spectrum is quasi degenerate.
We present this scenario, elaborate on its consequences and discuss its
realization.
\newpage
The vastly different mixing patterns \cite{rev} of quarks and leptons have been
used as an argument in favour of special  leptonic symmetries such as
$\mu$-$\tau$ interchange \cite{mutau1, mutau2,mutau3}, $L_e-L_\mu-L_\tau$
\cite{emt} , $D_4$ \cite{d4} ,
$A_4$ \cite{a4} {\it etc.}.
These symmetries lead to large or maximal mixing angles seen in the
leptonic sector. Logically, such symmetries would then not be
present in the quark sector which exhibits small mixing angles.
This need not be so and it is possible to describe both the quark and leptonic
mixing as a consequence of an approximately broken 2-3 symmetry
which exchanges the second and the third generation fermionic fields.
We argue that this symmetry  manifests
itself
more forcefully in the quark sector than in the leptonic sector and results in the
understanding of small values of $V_{ub}$ and $V_{cb}$ when it is broken at few percent level.
The leptonic mass matrices can also be thought to be nearly invariant under the 2-3 symmetry if the neutrino mass spectrum is quasi degenerate.

Let us first elaborate on the well-known \cite{mutau1,mutau2}
consequences of the $\mu$-$\tau$ symmetry. The light neutrino mass
matrix $\mnu$ is restricted to have the following form in the
presence of this symmetry:
\be
\label{mt} \mnu=\left( \ba{ccc}
X_\nu&A_\nu&A_\nu\\
A_\nu&B_\nu&C_\nu\\ A_\nu&C_\nu&B_\nu\\ \ea \right)~. \ee
This form leads to a maximal atmospheric mixing and zero $U_{e3}$
if it is assumed to be true in the flavour basis. In the same basis, the charged lepton mass
matrix is diagonal and consequently, it is not invariant under the
$\mu$-$\tau$ symmetry which would have implied $m_\mu=m_\tau$. It is possible to imagine
a larger symmetry ($e.g.~D_4$ \cite{d4}) which when broken leads to the above
form for $\mnu$ in the flavour basis.
In this case, the $\mu$-$\tau$ symmetry appears to be only an effective
neutrino symmetry.

It is important to stress that the $\mu$-$\tau$ symmetry by itself does not
force equality of the muon and tau
masses.
To see this, let us simultaneously assume that both the
charged lepton mass matrix $M_l$
and $\mnu$ are $\mu$-$\tau$ symmetric and have the form\footnote{The
2-3 symmetry does not automatically
imply the form given in in eq. (\ref{mt}) for $M_l$ unless it is
assumed to be symmetric. This assumption can  easily be realized
in the context of GUT such as $SO(10)$ which commutes with the 2-3
symmetry.} given in
eq. (\ref{mt}). In this
case, the muon and tau masses are different but now the
$23$ mixing angle for the charged leptons  is also maximal. As a
consequence, the neutrino and the charged lepton mixing angles cancel
and one gets vanishing atmospheric mixing angle. In either case, the
$\mu$-$\tau$ symmetry does not appear to be an exact symmetry in the leptonic world.

In contrast to  leptons, the $23$ and the $13$ mixing angles are indeed small for
quarks.
This suggests that a generalized $\mu$-$\tau$ symmetry may be a good
symmetry for quarks
rather than for leptons. Let us then postulate
that the quark  mass matrices are symmetric and display an approximate
2-3 symmetry. Latter on we will show that this assumption can be extended to the
leptonic masses as well.
An approximate 2-3 symmetry
dictates the following form for a symmetric fermion mass matrix
$M_{f}$:

\be
\label{mf} M_f=\left( \ba{ccc}
X_f&A_f(1-\e_{1f})&A_f(1+\e_{1f})\\
A_f(1-\e_{1f})&B_f(1-\e_{2f})&C_f\\ A_f(1+\e_{1f})&C_f&B_f(1+\e_{2f})\\
\ea \right)~. \ee
The dimensionless parameters $\e_{1f,2f}$ break the 2-3 symmetry and
are assumed to be $\ll 1$. These two parameters are sufficient
to describe the most general 2-3 breaking \cite{mutau2} when fermion mass
matrices are symmetric.

Let us first consider the symmetric limit assuming all parameters in eq. (\ref{mf}) to be real. All the eigenvalues of $M_f$ are distinct
and are given by
\beqa \label{ev}
m_{1f}&=& \frac{1}{2}\left[B_f+C_f+X_f-\left((B_f+C_f-X_f)^2+8 A_f^2\right)^{1/2}\right]~,\nonumber \\
m_{2f}&=& \frac{1}{2}\left[B_f+C_f+X_f+\left((B_f+C_f-X_f)^2+8 A_f^2\right)^{1/2}\right]~, \nonumber \\
m_{3f}&=& B_f-C_f ~.\eeqa
We will assume the hierarchy $|m_{1f}|<|m_{2f}|<|m_{3f}|$ and
associate the fermionic states accordingly to these eigenvalues.
The $M_{f}$ can be diagonalized by a matrix $V_f^0$:
\be V_{f}^0=R_{23}(\pi/4)R_{12}(\theta_{12f}) ~. \ee
As a result, one gets in the symmetric limit,
\be \label{vckm0}
V^0_{CKM}=V_u^{0\dagger}V_d^0=R_{12}({\theta_c})~,
\ee
with $$\theta_c\approx\theta_{12d}-\theta_{12u}.$$ It follows from
eq. (\ref{vckm0}) that the  2-3 symmetry automatically leads to
vanishing $V_{cb}$ and $V_{ub}$. This remains true even if $M_f$
is complex. The Cabibbo angle and the quark masses are not restricted
by this symmetry. The Cabibbo angle  can be constrained by
imposing an additional discrete symmetry $D$ defined as:
\be
\label{z4} f_{1L}\rightarrow i f_{1L}~~~~;~~~~ f_{1R}\rightarrow
-i f_{1R}~ . \ee 
This symmetry forces $A_f$ and $X_{f}$ in eq.
(\ref{mf}) to be zero. The $A_f$ term breaks this symmetry by one
and $X_f$ by two units (of $i$). $B_f$ and $C_f$ are invariant. It
is thus natural to assume that $D$-breaking (by some flavon field)
can lead to a hierarchy $|B_{f},C_{f}|>>|A_{f}|>>|X_f|$. This
hierarchy leads to $A_f\sim \ord(\sqrt{m_{1f}m_{2f}})$ and the
celebrated relation
\be \label{tc} \theta_c\sim  \sqrt{\frac{m_d}{m_s}}-
\sqrt{ \frac{m_u}{m_c}}~. \ee
More precisely, one needs,
\be \label{hier}
|X_f|\ll |\sqrt{2}A_f|\ll |B_f+C_f| \ll |B_f-C_f|~, \ee
for $f=u,d$ in order to get eq. (\ref{tc}) and the hierarchical masses.
It follows that an approximately broken $D$ and an exact 2-3 symmetry
leads to eq. (\ref{tc}) and vanishing $V_{ub},V_{cb}$. Subsequent breaking
of the 2-3 symmetry can then induce the latter quantities.

While both $\e_{1f}$ and $\e_{2f}$ could be present in a model, we
consider here one parameter breaking for all $M_f$ and assume that only
$\e_{2f}$ is non-zero. It is straightforward to add the effect of
$\e_{1f}$. We will also take all parameters to be real.

The non-zero $\e_{2u},\e_{2d}$ are sufficient to generate the
required values of $V_{ub}$ and $V_{cb}$. The $M_f$ can be
diagonalized in the limit specified in eq. (\ref{hier}) as follows
$$V_f^T M_f V_f={\rm Diag.}(m_{1f},m_{2f},m_{3f}),$$ with \beqa
\label{quarkm} m_{3f}\approx
~B_f-C_f(1+\frac{1}{2}\theta_{23f}^2),\nonumber \\ m_{2f}\approx
~B_f+C_f(1+\frac{1}{2}\theta_{23f}^2)+\frac{2
A_f^2}{m_{2f}},\nonumber \\ m_{1f}\approx ~ -\frac{2
A_f^2}{m_{2f}}, \eeqa 
where $f=u,d$. The mixing matrix is given as
\be  \label{vf} V_f=R_{23}(\pi/4) R_{23}(\theta_{23f})
R_{13}(\theta_{13f})R_{12}(\theta_{12f})~,\ee with 
\beqa
\label{quarkmixing} \theta_{23f}&\approx&
~\frac{\e_{2f}B_f}{2C_f}\approx -\frac{\e_{2f}}{2},\nonumber \\
\theta_{12f}&\approx& ~\sqrt{\frac{-m_{1f}}{m_{2f}}},\nonumber \\
\theta_{13f}&\approx&
~\frac{m_{2f}}{m_{3f}}\theta_{12f}\theta_{23f}. \eeqa
This leads to
 \beqa \label{ckm} V_{cb}&\approx&
~\theta_{23d}-\theta_{23u},\nonumber \\ 
V_{ub}&\approx&
\theta_{13d}-\theta_{13u}+\theta_{12u}(\theta_{23d}-\theta_{23u})\sim
\theta_{12u} V_{cb} \eeqa and eq. (\ref{tc}) for $V_{us}$.
Keeping a grand unified picture in mind, we assume that the $M_f$
in eq. (\ref{mf}) is defined at $M_{GUT}\sim 10^{16}$ GeV and
require it to reproduce the parameters in quark sector at that
scale. For definiteness, we choose the MSSM and quark masses
corresponding to $\tan\beta=10$ given in \cite{parida}.

It follows from eq. (\ref{ckm}) that a few percent breaking of the
2-3 symmetry can reproduce the observed mixing quite well for
several choices of parameters in $M_f$. For illustration, we give
one specific choice which is a typical phenomenologically
consistent example:
$$ \e_{2u}=-\e_{2d}\sim 0.045~, $$
 \be \label{ex1} \ba{cc}
M_d=\left( \ba{ccc} -0.003&0.0054&0.0054\\ 0.0054&0.49&-0.54\\
0.0054&-0.54&0.54\\ \ea \right)~&;~ M_u=\left( \ba{ccc}
0&0.0084&0.0084\\ 0.0084&42.74&-41.06\\ 0.0084&-41.06&39.055\\ \ea
\right)~. \ea \ee

These mass matrices lead to the mixing angles $|V_{us}|\approx
0.221$, $|V_{cb}|\approx 0.044$ and $|V_{ub}|\approx 0.0026$.
These values are in approximate agreement  with the high scale
estimates $|V_{us}|\sim 0.223-0.226$, $|V_{cb}|\sim 0.029-0.038 $
and $V_{ub}\sim 0.0024-0.0038$ as given for example in Matsuda and
Nishiura, ref. \cite{mutau3}. This agreement can be improved by
switching on small $\e_1$. The approximate 2-3 symmetry of quark
mass matrices is apparent in eq. (\ref{ex1}).

Let us now turn to the leptonic sector. Our discussion of the quark
sector shows that the smallness of $V_{cb}$  is
a natural consequence of the 2-3 symmetry. The corresponding mixing angle
for leptons is known to be almost maximal. This has led to a view \cite{mutau1,mutau2} that the $\mu$-$\tau$ symmetry is an effective symmetry of neutrino mass
matrix  badly broken in the  charged lepton sector.
This need not always be the case as we argue now.

We assume that just as in case of the quarks, both $M_l$ and
$\mnu$ are  having approximate 2-3 symmetric forms given in eq.
(\ref{mf}). The $\mnu$ here refers to the effective mass matrix of
the light neutrinos. It can originate from an approximate  2-3
symmetric couplings to a triplet Higgs or may originate from the
ordinary seesaw mechanism in which the Dirac neutrino mass matrix
$m_D$ and the right handed neutrino mass matrix $M_R$ are
approximately 2-3 symmetric with the form given in eq. (\ref{mf}).

Assume that the $A_{\nu,~l}$ are small parameters as in case of the quarks
and concentrate first on the lower $2\times 2$ block of eq. (\ref{mf}). Its diagonalization
gives
\be \label{enu}
\e_{2f}= \left( \frac{m_{2f}-m_{3f}}{m_{2f}+m_{3f}}\right) \cos 2
\tilde{\theta}_{23f} ~. \ee
$f=l,\nu$ above and $\tan 2\tilde{\theta}_{23f}\equiv
\frac{C_f}{\e_{2f}B_f}$ correspond to the $23$ mixing angle for
$f$. This equation gives a clue to obtaining approximate $23$
symmetry simultaneously for $M_l$ and $M_\nu$ as well as  large
atmospheric mixing angle. The approximate 2-3 symmetry requires
$\e_{2\nu},\e_{2l}\ll 1$. For the charged leptons, small $\e_{2l}$
necessarily means $\tilde{\theta}_{23l}\sim \frac{\pi}{4}$ in eq.
(\ref{enu}) since $m_\mu$ substantially differs from $m_{\tau}$.
In contrast, for neutrinos small  $\e_{2\nu}$ can be realized
either with a large $\tilde{\theta}_{23\nu}$ or with  $m_{2\nu}\sim
m_{3\nu}$. The latter case will correspond to a large atmospheric
mixing angle. It follows that in case of the quasi degeneracy, there
exists ranges in parameters corresponding to approximately 23
symmetric $M_{l}$ and $\mnu$ and large atmospheric mixing arising
due to a small $\tilde{\theta}_{23\nu}$ and almost maximal $\tilde{\theta}_{23l}$.
All three neutrinos are required to be quasi degenerate in order
to obtain simultaneous explanation of the solar and atmospheric
neutrino scales. In particular, the $m_{\nu_2}$ and $m_{\nu_3}$
would need to have the same sign to make $\e_{2\nu}$ small.

The 2-3 symmetry can be exact in $M_l$ while it needs to be broken
by $\mnu$. The amount of the required breaking is quantified using
eq. (\ref{enu}): \be \label{e2nu} |\e_{2\nu}|\approx
\left|\frac{m_{\nu_2}-m_{\nu_3}}{m_{\nu_2}+m_{\nu_3}}\right|\approx
\left|\frac{\Delta_A}{4 m_0^2}\right|\sim 0.08 \ee for the
atmospheric scale $$\Delta_A\sim 3\times 10^{-3}~\eV^2$$ and the
quasi degenerate mass $m_0\sim 0.1 ~\eV$. This value is not very
different from the symmetry breaking that was required in the
quark sector.

In order to analyze the leptonic mixing in the full $3\times 3$ case,
let us assume that $M_l$ is 2-3 symmetric and go to the basis with a
diagonal $M_l$. In this basis, the neutrino
mass matrix assumes the form
\be \label{mnuf}
{\cal M}_{\nu f}\equiv R_{12}^T(\theta_{12l})R_{23}^T(\pi/4) \mnu
R_{23}(\pi/4) R_{12}(\theta_{12l})~. \ee
The $\theta_{12l}$ denotes the $e$-$\mu$ mixing which in analogy with
the quark case
will be assumed to be small, $\theta_{12l}\sim
\sqrt{\frac{m_e}{m_{\mu}}}$.
Neglecting its effect, the ${\cal M}_{\nu f}$ is approximately given
by
\be \label{mnuf2}{\cal M}_{\nu f}\approx \left( \ba{ccc}
 X_\nu&\sqrt{2}A_\nu&0\\
\sqrt{2}A_\nu&B_\nu+C_\nu&\e_{2\nu}B_\nu\\
0&\e_{2\nu}B_\nu&B_\nu-C_\nu\\ \ea \right) ~.\ee ${\cal M}_{\nu
f}$ is diagonalized by the PMNS matrix \cite{pmns} $U$ as 
\be
\label{pmns} U^T {\cal M}_{\nu f} U={\rm
Dia.}(m_{\nu_1},{m_{\nu_2},m_{\nu_3}}) ~, \ee 
with $U=R_{23}(\theta_{23})
R_{13}(\theta_{13}) R_{12}(\theta_{12})$ in the standard parameterization.

Consider the symmetric limit corresponding to $\e_{2\nu}=0$. The quasi degeneracy
$m_{\nu_2}\sim m_{\nu_3}$ is obtained for
\be \label{qdg}
B_\nu\sim m_0~~~;~~~C_\nu\sim \ord\left(\frac{\Delta_{A}}{4 m_0}\right)~. \ee
The atmospheric mixing is zero in this case but when $\e_{2\nu}$ is turned on, even a small
value as given in eq. (\ref{e2nu})  can lead to a large atmospheric mixing due to smallness of $C_\nu$
The smallness of $C_\nu$, $i.e.$ the quasi degeneracy does not follow from the underlying 2-3 symmetry but it is
quite consistent with it.

The expression for the atmospheric mixing angle
follows from the diagonalization of
the  $23$ block
\be \label{atm}
\tan 2 \theta_{23}=\frac{\e_{2\nu}B_\nu}{C_{\nu}}~. \ee
This gets a small correction when $A_\nu  \sim \ord
(\frac{\Delta_{\odot}}{4 m_0})$ is turned on.

While the small 2-3 breaking leads to a large atmospheric mixing, the $U_{e3}$ remains small. This follows because of the zero in
eq. (\ref{mnuf2}) at the (13) entry. Using
$({\cal M}_{\nu f})_{13}=(U D_\nu U^T)_{13}=0$ and the quasi degeneracy, one finds
\be \label{ue3} U_{e3}\sim \tan\theta_{23}\sin2 \theta_{12}\frac{\Delta_\odot}{2\Delta_A}\sim \pm 0.03~,\ee
where $\Delta_A\equiv m_{\nu_3}^2-m_{\nu_1}^2$ and $\Delta_\odot\equiv m_{\nu_2}^2-m_{\nu_1}^2$. Note that the normal and inverted neutrino mass hierarchies correspond to opposite signs for $U_{e3}$.

The above value for $U_{e3}$ would get corrected by ($a$)  the 12
mixing angle in the charged lepton sector and ($b$) the symmetry
breaking parameter $\e_{1\nu}$ which was also neglected here. The
($a$) gives a contribution \cite{ue3} of
$$\ord\left(\frac{1}{\sqrt{2}}\theta_{12l}\right)\sim 0.05.$$
which can add or subtract to the value $\sim 0.03$ given above
depending upon the neutrino mass hierarchy. There can be a
relative phase between these contribution in the presence of CP
violation. As a consequence, one expects $U_{e3}$ in the present
scheme to be typically $0.02- 0.08$ if $\theta_{12l}\sim
\ord\left(\frac{1}{\sqrt{2}}\theta_{12l}\right)$. The $\e_{1\nu}$
gives a very small $\sim
\ord(\frac{\Delta_\odot}{\Delta_A}\e_{1\nu})$ contribution to
$U_{e3}$ when $A_\nu\sim \ord(\frac{\Delta_\odot}{m_0})$.

\begin{figure}[h]
\centerline{\psfig{figure=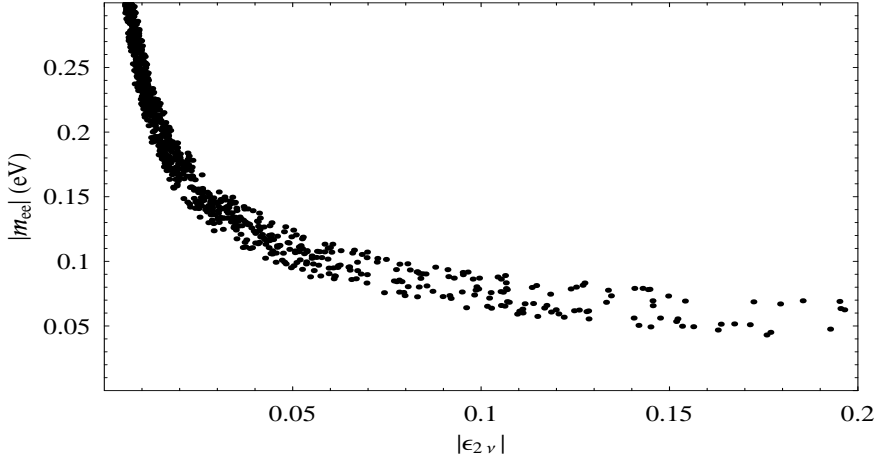,height=10cm,width=12cm,angle=0}}
\vskip  0.5cm \caption{The allowed ranges of the 23-breaking
parameter $\e_{2\nu}$ and the neutrinoless double beta decay mass
$m_{ee}$ obtained from eq.(\ref{mnuf2}) with quasi degenerate
spectrum. The solar and the atmospheric scales and mixing angles
are  randomly varied within their allowed 2$\sigma$ ranges. }
\end{figure}
\vskip 0.5cm

The quasi-degeneracy is an essential ingredient in this approach.
One would therefore expect relatively large value for the
effective neutrino mass $m_{ee}$ probed by the neutrinoless double
beta decay experiments. This is quantified in  in Figure 1. The
parameters in $M_{\nu_f}$ are determined in terms of the lightest
neutrino mass $m_0$, the solar and the atmospheric scales  and the
corresponding  mixing angles  using eq. (\ref{mnuf2},\ref{pmns})
after imposing eq. (\ref{ue3}). These are then varied randomly in
their allowed $2\sigma$ ranges \cite{fogli} to generate the values
of $m_{ee}$ and $\e_{2 \nu}$. The sum of neutrino masses is
assumed to be $\leq 0.9 ~\eV$ as required by cosmology. One
clearly sees that quite large values for $m_{ee}$ are possible
which is understood from the fact that the scenario corresponds to
quasi degeneracy with all the neutrinos having the same CP
property. The $\e_{2\nu}$ is restricted in the range $\sim
0.005-0.2$ with higher $m_0$ requiring smaller 23 breaking.

The atmospheric mixing angle can be large but it is not
required to be maximal as would be the case if
only ${\cal M}_{\nu f}$ was assumed to be $\mu$-$\tau$ symmetric.
While strict maximality does not obtain, all values allowed by the
present data are possible including close to the maximal
mixing.

We now turn to a concrete realization of our basic ansatz $M_f$
given in eq. (\ref{mf}). This can be derived in a straightforward
manner within the standard  two double model by imposing a 2-3
symmetry on the Yukawa couplings. One of the doublets ($\phi_1$)
is assumed to be invariant while the other ($\phi_2$) is odd under
the 2-3 symmetry. The Yukawa couplings for a fermion $f$ are then
given by 
\be \label{yukawa} -{\cal L}_Y=\bar{f}_L
(\Gamma_1\phi_1+\Gamma_2\phi_2) f_R~+~ {\rm H.c.} ~. \ee
The (assumed) symmetry of  $\Gamma_{1,2}$ and the 2-3 symmetry
together lead to
the matrix $M_f$. The $\Gamma_1$ generates
the parameters $A_f,B_f,C_f$ in eq. (\ref{mf}) and  $\Gamma_2$ 
generates $\e_{1f,2f }$ terms. The
smallness of $\e_{1f,2f }$ compared to the leading elements can be
obtained by assuming corresponding elements of $\Gamma_{1,2}$ to
be similar but taking
$\frac{\langle\phi_2\rangle}{\langle\phi_1\rangle}$ to be small
$\leq 0.1$

The neutrino mass matrix also follows in a straightforward manner, $e.~g.$
consider a model with right handed neutrinos which obtain mass from a standard model singlet (or 126 in $SO(10)$) assumed to be invariant under the 2-3 symmetry.
The mass matrix $M_R$ would then be 2-3 symmetric. This together with the
Dirac mass matrix obtained from eq. (\ref{yukawa}) would lead to
a neutrino mass matrix having the form of eq. (\ref{mf}).

Simplicity of the above scheme is to be contrasted with other
models \cite{d4} which try to obtain a $\mu$-$\tau$ symmetric
neutrino mass matrix in the flavour basis.

In summary, let us recapitulate the salient features of the scheme
and open problems.
\begin{itemize}
\item We showed that the $\mu$-$\tau$ symmetry can be extended to all
fermions with interesting consequences. Many earlier studies
\cite{mutau1,mutau2} postulated this
only for the neutrino mass matrix in flavour basis and its extension
to other fermions was found problematic. As shown here, approximate
23 symmetry is quite consistent with observations if neutrinos
have quasi degenerate spectrum. We quantified the amount of
breaking of the 2-3 symmetry needed for successful phenomenology.
\item In the quark sector, the 2-3 symmetry provides explanation of the smallness
of $V_{cb},V_{ub}$ compared to the Cabibbo angle. The latter can
be naturally explained if an additional symmetry $D$ as defined in
eq. (\ref{z4}) is imposed. This  needs to be broken badly by the
effective neutrino mass matrix in order to get the
quasi-degenerate spectrum.
\item The $\mu$-$\tau$ symmetry has been extended to the quark sector in some of the earlier works \cite{mutau3}. These relied on breaking it  through complex phases in the mass matrix. In contrast, the 2-3 symmetry breaking here occurs even when the phases are turned off but requires quasi degenerate neutrino spectrum. This feature can be tested through the neutrinoless double beta decay and direct neutrino mass  measurements in future experiments .
\item The maximal atmospheric mixing is one of the predictions of the
unbroken $\mu$-$\tau$ symmetry of the neutrino mass matrix. This
maximality is not obtained here but values very close to the maximal
are possible.  While realization of the effective $\mu$-$\tau$
symmetry for neutrino requires complicated models \cite{d4}, the
present scenario gets realized in the standard two Higgs doublet
model.
\item The Yukawa couplings in eq. (\ref{yukawa}) generate the flavour
changing neutral currents (FCNC). One finds that the specific
structures of the Yukawa couplings $\Gamma_{1,2}$ lead to
hierarchical strengths ($|F_{12}|\ll |F_{13}|\ll |F_{23}|)$ for
the FCNC current couplings $F_{ij}$  between flavours $i$ and $j$
to Higgs in case with one parameter symmetry breaking, $i.e.$,
with $\e_2\not=0$. Rough estimates give for the  down quarks,
$F_{12}\sim \frac{m_b}{v} \lambda^5,F_{13}\sim \frac{m_b}{v}
\lambda^3$ and $F_{23}\sim \frac{m_b}{v} \lambda^2$, where
$\lambda\sim 0.22$ and $v$ is the weak scale. Because of this
suppression, the FCNC do not require unusually large Higgs mass.
Similar hierarchy in FCNC is found in some Higgs doublets models
\cite{fch} and in $Z$ couplings \cite{fcz} in models with
additional quarks.
\item As in  case of the $\mu$-$\tau$
symmetry, $U_{e3}$
remains small but can be close to the measurable values in future.
\item We assumed CP conservation and identical CP properties for the
neutrinos. The latter is required to make symmetry breaking
parameter small.
\item The entire scenario is compatible with grand unification and can be
embedded in theories such as SO(10).
\end{itemize}

\end{document}